\documentclass[%
aip,apl,reprint
]{revtex4-1}
\usepackage{eqnarray,amsmath}
\usepackage{graphicx}
\usepackage{dcolumn}
\usepackage{bm}
\usepackage{wasysym}
\usepackage{gensymb}
\usepackage{color}
\definecolor{amber}{rgb}{1.0, 0.49, 0.0}
\usepackage[normalem]{ulem}
\usepackage{calrsfs}
\usepackage{float}
\DeclareMathAlphabet{\pazocal}{OMS}{zplm}{m}{n}
\newcommand{\Ca}{\mathcal{C}}

\newcommand{\abs}[1]{\left\lvert#1\right\rvert}
\definecolor{light-gray}{gray}{0.45}
\definecolor{deeogreen}{RGB}{34,159,34}
\begin{document}

\title{Miniature cavity-enhanced diamond magnetometer}

\author{Georgios Chatzidrosos}
\email{gechatzi@uni-mainz.de}
\affiliation{Johannes Gutenberg-Universit{\"a}t  Mainz, 55128 Mainz, Germany}
\author{Arne Wickenbrock}
\affiliation{Johannes Gutenberg-Universit{\"a}t  Mainz, 55128 Mainz, Germany}
\author{Lykourgos Bougas}
\affiliation{Johannes Gutenberg-Universit{\"a}t  Mainz, 55128 Mainz, Germany}
\author{Nathan Leefer}
\affiliation{Johannes Gutenberg-Universit{\"a}t  Mainz, 55128 Mainz, Germany}
\author{Teng Wu}
\affiliation{Johannes Gutenberg-Universit{\"a}t  Mainz, 55128 Mainz, Germany}
\author{Kasper Jensen}
\affiliation{Niels Bohr Institute, University of Copenhagen, Blegdamsvej 17, 2100 Copenhagen, Denmark}
\author{Yannick Dumeige}
\affiliation{CNRS, UMR 6082 FOTON, Enssat, 6 rue de Kerampont, CS 80518, 22305 Lannion cedex, France}
\author{Dmitry Budker}
\affiliation{Johannes Gutenberg-Universit{\"a}t  Mainz, 55128 Mainz, Germany}
\affiliation{Helmholtz Institut Mainz, 55099 Mainz, Germany}
\affiliation{Department of Physics, University of California, Berkeley, CA 94720-7300, USA}
\affiliation{Nuclear Science Division, Lawrence Berkeley National Laboratory, Berkeley, CA 94720, USA}

\date{\today}

\begin{abstract}

We present a highly sensitive miniaturized cavity-enhanced room-temperature magnetic-field sensor based on nitrogen-vacancy (NV) centers in diamond. The magnetic resonance signal is detected by probing absorption on the 1042\,nm spin-singlet transition. 
To improve the absorptive signal the diamond is placed in an optical resonator.
The device has a magnetic-field sensitivity of 28\,pT/$\sqrt{\rm{Hz}}$, a projected photon shot-noise-limited sensitivity of 22\,pT/$\sqrt{\rm{Hz}}$  and an estimated quantum projection-noise-limited sensitivity of 0.43\,pT/$\sqrt{\rm{Hz}}$ with the sensing volume of $\sim$\,390\,$\mu$m $\times$\,4500\,$\mu$m$^{2}$. The presented miniaturized device is the basis for an endoscopic magnetic field sensor for biomedical applications.

\end{abstract}

\maketitle

\section*{Introduction}
Biomagnetic signatures are an important diagnostic tool to understand the underlying biological processes.
Time-resolved biomagnetic signals are measured with Hall probes\citep{Manandhar2009}, Giant magnetoresistance sensors\citep{Barbieri2016}, alkali-vapor magnetometers\cite{Jensen2016scie}, superconducting quantum interference devices (SQUIDs)\cite{Ulusar2011} and single negatively-charged nitrogen-vacancy (NV) centers or ensembles thereof\cite{Barry2016}.
Typical devices probe magnetic fields outside the body, i.e., far from their origin. 
However, signal strength and spatial resolution can both be improved by utilizing endoscopic sensors.

NV centers in diamond have already been used as nanoscale-resolution sensors\,\cite{Balasubramanian2008,Maze2008,Rittweger2009} with high sensitivity\,\cite{Barry2016,Wolf2015}.
Prominent examples of sensing with NV centers include, single neuron-action potential detection\,\cite{Barry2016}, single protein spectroscopy\,\cite{Lovchinsky2016}, as well as in vivo thermometry\,\cite{SingleNVThermometry}.
Due to their ability to operate in a wide temperature range as well as their small size, NV magnetometers are amenable for in-vivo and/or endoscopic applications.

The majority of NV sensors use a photoluminescence (PL) detection which suffers from low photon-detection efficiency.
Approaches to counter this problem include, for example, the use of  solid immersion lenses\cite{Hadden2010,Siyushev2010,Sage2012}, or employ infrared (IR) absorption\cite{Dumeige2013,Budker1,Jensen2014}.
Compared to NV sensors based on PL detection, those based on absorption feature collection efficiency approaching unity\cite{Budker1}.

Due to the small cross-section of the IR transition\,\cite{Budker1,Dumeige2013}, to achieve similar or higher sensitivities compared to PL-detection techniques, we use an optical cavity to enhance the optical pathlength in the diamond, and thus the IR absorption signal\,\cite{Dumeige2013,Jensen2014}.
With the cavity enhancement we can achieve sensitivities closer to the fundamental projection-noise limit, even at room temperature\cite{Dumeige2013,Jensen2014}. Here we demonstrate a sensitive compact cavity-based IR absorption device operating near the photon shot-noise limit opening realistic prospects for a practical endoscopic magnetometer. 

\begin{figure*}[t]
\centering
\includegraphics[width=\textwidth]{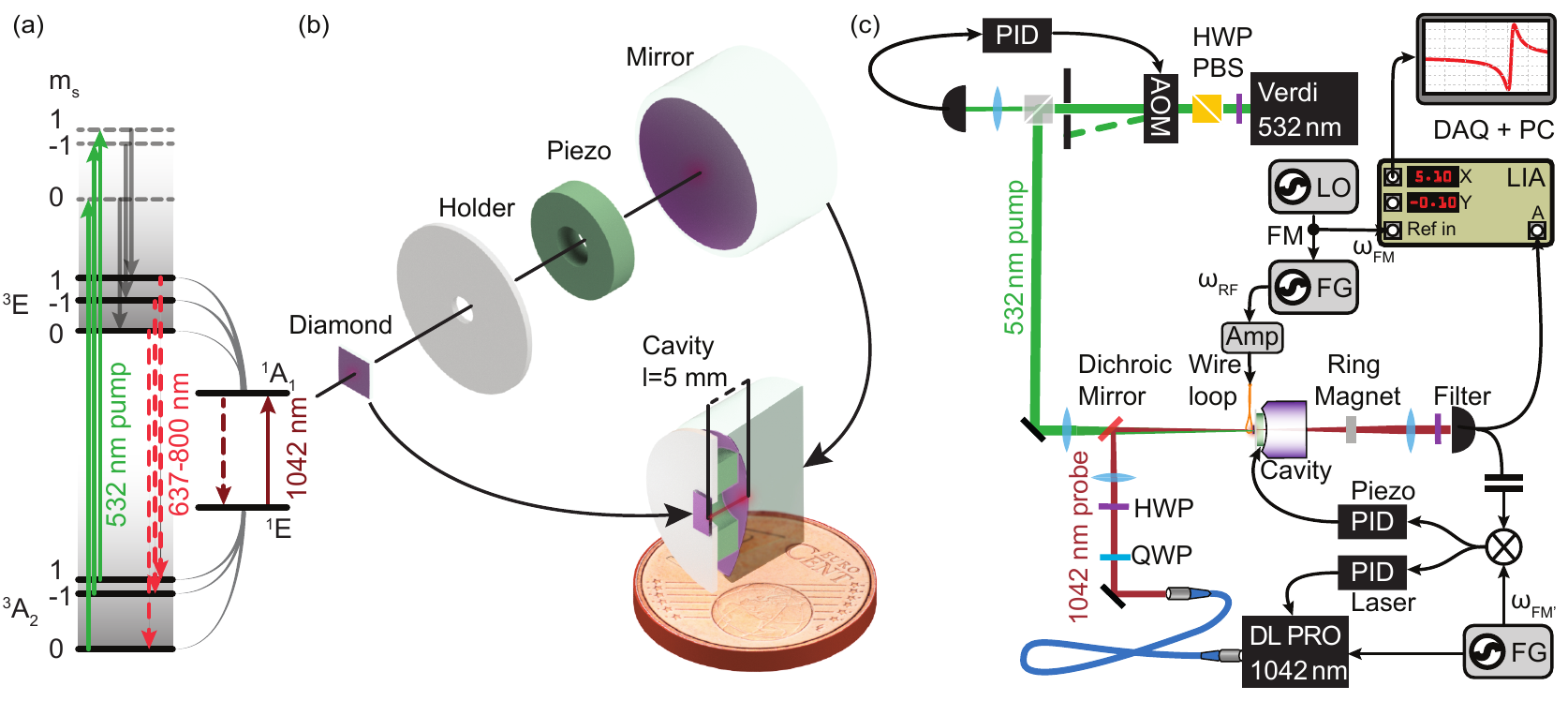}
\caption{\small{(a) Relevant NV center energy levels and transitions. Solid green and red lines indicate excitations, dashed lines indicate radiative transitions, and gray solid lines indicate non-radiative transitions.
(b) Explosion view of the cavity magnetometer.
(c) Schematic of the experimental setup.}}
\label{fig:fig1}
\end{figure*}
\section*{Experiment}
The ground and excited electronic spin-triplet states of NV are $^{3}$A$_{2}$ and $^{3}$E, respectively [Fig.\,\ref{fig:fig1}\,(a)], with the transition between them having a zero-phonon line at 637\,nm. 
The lower and upper electronic singlet states are $^{1}$E and $^{1}$A$_{1}$, respectively, with the transition between them having a zero-phonon line at 1042\,nm (IR). 
While optical transition rates are spin-independent, the probability of nonradiative intersystem crossing from $^3$E to the singlets is several times higher\cite{Wrachtrup2016} for $m_s=\pm 1$ than that for $m_s=0$. 
As a consequence, under continuous illumination with green pump light (532\,nm), NV centers are prepared in the $^{3}$A$_{2}$ $m_s=0$ ground state sublevel and in the metastable $^{1}$E singlet state. 
For metrology applications, the spins in the the $^{3}$A$_{2}$ ground state can be coherently manipulated by microwave fields. 
In this work, the population of the ground state is inferred by monitoring IR light absorption on the singlet transition.\\

\indent To increase the absorption of IR light we construct a cavity
as shown in Fig.\,\ref{fig:fig1}\,(b).
A spherical mirror with a curvature radius of 10\,mm, reflectivity of 99.2(8)\% (see supplemental material\cite{Supplemental}) and diameter of 12.5\,mm serves as the output coupler.
A piezoelectric transducer is used to adjust the length of the cavity within a range of a few $\mu$m. It is glued with an epoxy resin (Torr Seal) between the spherical mirror and the ceramic holder for the diamond. The diamond plate serves as the input plane mirror of the cavity and is glued to the holder. The holder doubles as a heat sink.
The (111)-cut diamond plate is dielectrically coated with high reflectivity $\sim$\,98.5\% for IR light as well as anti-reflective for green light on the outside of the cavity. 
The diamond surface inside the cavity is supplied with an anti-reflective coating for both green light and IR light.
The total optical length of the cavity is 5.00(3)\,mm, and the finesse is $\mathcal{F}$\,=\,160(4) (see supplemental material\cite{Supplemental}). 
The cavity mode has a waist on the diamond with a radius of $38\,\mu$m; the mode radius is 54$\,\mu$m on the concave mirror surface.
With this design, it is possible to bring the diamond's outer surface in close proximity to a magnetic sample under study in compact geometry.\\
\indent The setup for magnetometric measurements is shown in Fig.\,\ref{fig:fig1}\,(c). 
Green light is provided by a diode-pumped solid-state laser (Coherent, Verdi V10) and IR light is provided by an external-cavity diode laser (Toptica, DL-Pro).
The green laser power is stabilized using an acousto-optical modulator (AOM, ISOMET-1260C with an ISOMET 630C-350 driver) controlled through a proportional-integral-derivative controller (PID, SIM960). 
The IR beam profile is matched to the lowest-order longitudinal cavity mode (TEM$_{00}$), while the green beam is overlapped with the IR beam in the center of the cavity; it is not necessary to exactly mode-match the green beam profile. 
The frequency of the IR laser is locked to the cavity mode using a modulation technique with feedback to two PID controllers (SIM960).
Fast feedback is realized by changing the laser current, while the cavity piezo actuator is used for slow feedback.

The microwaves (MW) to manipulate the NV spins are generated by a MW generator (SRS SG394). They are amplified with a 16\,W amplifier (ZHL-16W-43+), passed through a circulator (CS-3.000, not shown in Fig.1) and high-pass filtered (Mini Circuits VHP-9R5), before they are applied to the NV centers using a mm-sized wire loop. The other side of the wire is directly connected to ground. A bias magnetic field is applied with a permanent ring magnet mounted on a precision positioning stage.\\

\begin{figure}[t]
\centering
\includegraphics[width=\columnwidth]{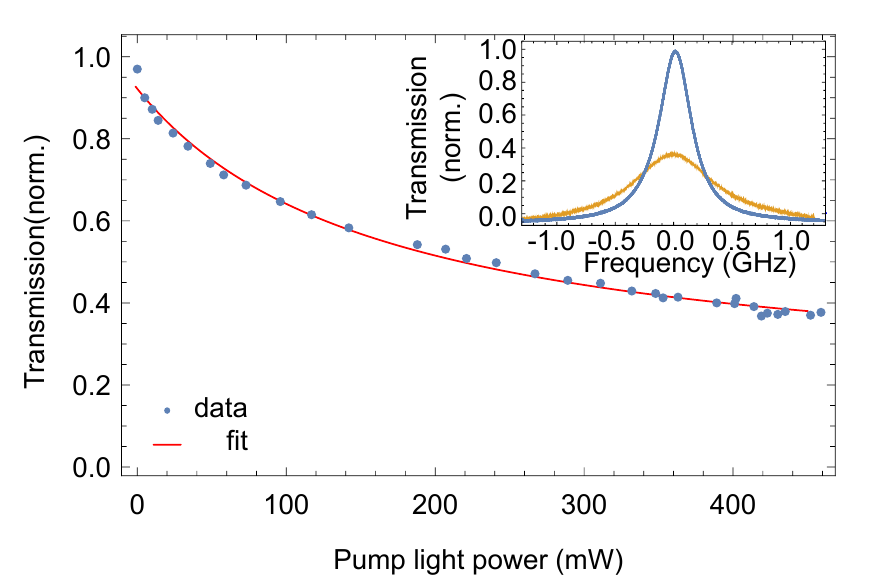}
\caption{\small{IR light transmission of the cavity as a function of pump power. The transmission is normalized to unity for zero pump power.
The inset displays the TEM$_{00}$ cavity mode at 0\,mW \color{blue}(blue) \color{black} and 500\,mW of pump light power \color{amber}(amber) \color{black}.}}
\label{fig:fig2}
\end{figure}

\section*{Results and discussion}
\begin{figure}[t]
\centering
\includegraphics[width=\columnwidth]{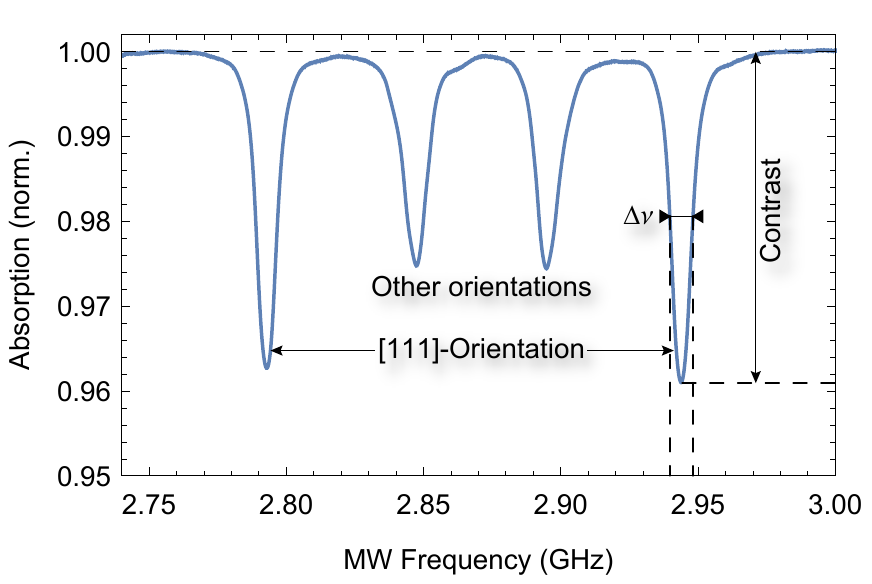}
\caption{\small{IR transmission signal normalized to unity off resonance as a function of microwave frequency.}} 
\label{fig:fig3}
\end{figure}

\indent The cavity transmission signal $I_\textrm{tr}$ for IR light is shown in Fig.\,\ref{fig:fig2} as a function of green light power $P$ in front of the cavity.
The steady-state population of the singlet state increases with increasing green power,  resulting in higher IR absorption. 
The IR absorption is enhanced by the cavity by 2$\mathcal{F}/\pi$, yielding significantly reduced IR transmission for higher pump powers. 
Higher absorption also results in an increase of the cavity-mode linewidth (Fig.\,\ref{fig:fig2}, inset).
The data in Fig.\,\ref{fig:fig2} are fitted with a saturation curve
$I_\textrm{tr}\,=\,1 -\alpha P/(P+P_\textrm{sat})$ (Ref.\cite{Jensen2014}), with saturation power $P_\textrm{sat}$ and reduction in transmission at saturation $\alpha$. The fit results are $P_\textrm{sat}$\,=\,735(1)\,mW and $\alpha$\,=\,0.605(1).
Magnetic-resonance measurements (Fig.\,\ref{fig:fig3}) are performed by scanning the MW frequency around the NV zero-field splitting (2.87\,GHz).
When the MW field is resonant with the ground-state $m_s=0$ $\rightarrow$ $m_s=\pm 1$ transitions, population is transferred through the exited triplet state to the metastable singlet state, resulting in increased IR absorption, which produces the observed optically detected magnetic resonance (ODMR) signal.

With a bias magnetic field (in this case, about 3\,mT) aligned along the [111] axis, 
four peaks are visible by scanning the MW frequency (Fig.\,\ref{fig:fig3}). 
The outer features result from the  NVs along the [111] axis and the inner features from the remaining NV orientations.
The contrast $\Ca$ and the full width at half maximum $\Delta\nu$ of the outer peaks are $\sim$ 3.7\,$\%$ and 5.6\,MHz, respectively.

\begin{figure}[t]
\centering
\includegraphics[width=\columnwidth]{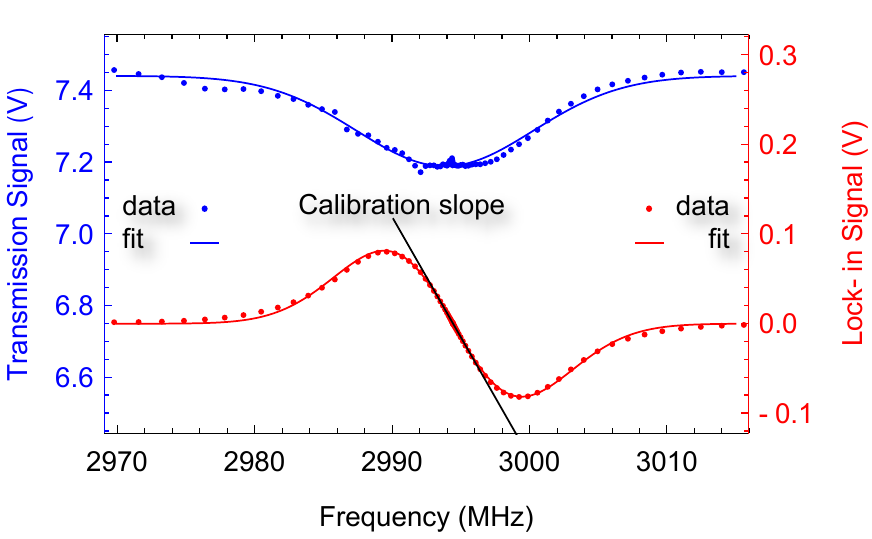}
\caption{\small{The cavity-transmission signal \color{blue} (blue) \color{black} and lock-in demodulated signal \color{red} (red) \color{black} as a function of the microwave frequency scanned over a magnetic resonance. The experimental data are represented with dots and the fits with solid lines. The slope of the fit is represented with a straight line.}}
\label{fig:fig4}
\end{figure}

\begin{figure}[t]
\centering
\includegraphics[width=\columnwidth]{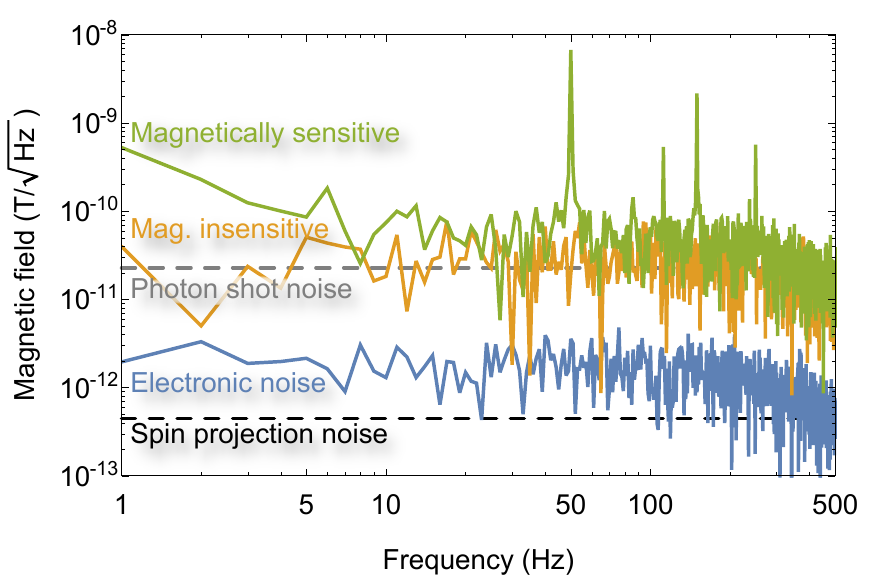}
\caption{\small{Magnetic-field noise spectrum \color{deeogreen} (green) \color{black} represents the magnetically sensitive spectrum corresponding to a environment limited noise floor of 37\,pT/$\sqrt{\rm{Hz}}$ between 60 and 90 Hz, \color{amber} (amber) \color{black} the magnetically insensitive \color{black} with a noise floor of 28\,pT/$\sqrt{\rm{Hz}}$ and \color{blue} (blue) \color{black}  the electronic noise with a floor of 2\,pT/$\sqrt{\rm{Hz}}$. The shot noise limit of the system and the spin projection noise are also depicted with a dashed lines at 22\,pT/$\sqrt{\rm{Hz}}$ and 0.43\,pT/$\sqrt{\rm{Hz}}$, respectively.}}
\label{fig:fig5}
\end{figure}

For the magnetometric measurements we focus on the highest frequency feature in Fig.\,\ref{fig:fig3}. We modulate the MW frequency $f_{\textrm{MW}}$ around the central frequency $f_{\textrm{c}}$ of the feature with frequency $f_{\textrm{mod}}$\,=\,8.6\,kHz and modulation amplitude $f_{\textrm{dev}}$\,=\,4.5\,MHz: $f_{\textrm{MW}} = f_{\textrm{c}} + f_{\textrm{dev}}\textrm{cos}(2 \pi f_{\textrm{mod}} t)$ and detect the first harmonic of the transmission signal with a lock-in amplifier (LIA).
Fig.\,\ref{fig:fig4} shows the resulting dispersive signal centered at the feature (red) along with the feature itself (blue).
Around the zero-crossing of the dispersive feature, we observe a linear signal S$_{\textrm{LI}}\sim\alpha(f_{\textrm{c}}-f_{\textrm{res}})$ as a function of $(f_{\textrm{c}}-f_{\textrm{res}})$ when $\abs{f_{\textrm{c}}-f_{\textrm{res}}}<<\Delta\nu/2$.
We extract the slope $\alpha$\,$\propto$\,$\Ca$/$\Delta\,\nu$ from the fit of Fig.\,\ref{fig:fig4} (black)
and use it to convert the magnetometer's voltage output into magnetic field.\\
Figure\,\ref{fig:fig5} shows the magnetic-field-noise spectrum. The spectrum was obtained by a Fourier transform of the LIA output with a reference frequency of 8.6\,kHz. 
The peaks at 50\,Hz and harmonics are attributed to magnetic field from the power line in the lab and are not visible on the magnetically insensitive spectrum, which we obtain in the absence of a MW field.
The noise floor in the region of 60-90\,Hz for the magnetically insensitive spectrum is calculated as 28\,pT/$\sqrt{\rm{Hz}}$. This sensitivity is $\approx$\,100 times better than what has been demonstrated previously with magnetometers based on IR absorption \cite{Jensen2014,Budker1}. Main improvements are: a dramatic reduction in cavity size, increase in probe laser power and improvements to the laser-lock stability. We verify the sensitivity by applying test magnetic fields (see supplemental material\cite{Supplemental}).
The photon shot noise limit is estimated as 22\,pT/$\sqrt{\rm{Hz}}$ for 4.2\,mW of collected IR light. 
The electronic shot noise is 2\,pT/$\sqrt{\rm{Hz}}$.
For an estimated NV density in the metastable singlet state of 0.68(1)\,ppm (see supplemental material\cite{Supplemental}) and the demonstrated ODMR linewidth $\Delta\nu = 5.6$\,MHz we calculate a spin-projection noise limit of 0.43\,pT/$\sqrt{\rm{Hz}}$.
The bandwidth of the magnetometer is set by the LIA filter settings. For the presented measurements a time constant of $300\,\mu$s results in a 530\,Hz bandwidth. The filter steepness is selected as 24dB/octave.

We demonstrate a miniaturized cavity-enhanced room-temperature absorption-based magnetometer using NV centers in diamond. 
The small size of our magnetometer yields a robust device with improved magnetic field sensitivity and makes it an ideal candidate for endoscopic measurements.
The closer proximity to biomagnetic signal sources, inherent to endoscopic measurements, provides enhanced signal strength and spatial resolution, which may be further improved by using a combination of a different diamond and a higher finesse cavity.
Our sensor features a noise-floor of 28\,pT/$\sqrt{\rm{Hz}}$ close to the shot-noise limit (see supplemental material\cite{Supplemental}). 
The sensitivity may be improved in future iterations by increasing the IR light power\cite{AcostaPRB2010}, using a critically matched cavity\cite{Dumeige2013}, implementing AC sensing protocols that allow increased NV coherence times due to dynamic decoupling from the decoherence sources\cite{Farfurnik2015}, and using a diamond sample with narrower linewidth. 
\begin{acknowledgements} 
We acknowledge support by the DFG through the DIP program (FO 703/2-1). GC acknowledges support by the internal funding of JGU. NL acknowledges support from a Marie Curie International Incoming Fellowship within the 7th European
Community Framework Programme. LB is supported by a Marie Curie Individual Fellowship within the second Horizon 2020 Work Programme. DB acknowledges support from the AFOSR/DARPA QuASAR program. We thank J.W. Blanchard for a fruitful discussion.
\end{acknowledgements}

\section*{References}
\bibliography{literature}

\end{document}


\beginsupplement

\title{Supplemental material to: Miniature cavity-enhanced diamond magnetometer}

\author{Georgios Chatzidrosos}
\email{gechatzi@uni-mainz.de}
\affiliation{Johannes Gutenberg-Universit{\"a}t  Mainz, 55128 Mainz, Germany}
\author{Arne Wickenbrock}
\affiliation{Johannes Gutenberg-Universit{\"a}t  Mainz, 55128 Mainz, Germany}
\author{Lykourgos Bougas}
\affiliation{Johannes Gutenberg-Universit{\"a}t  Mainz, 55128 Mainz, Germany}
\author{Nathan Leefer}
\affiliation{Johannes Gutenberg-Universit{\"a}t  Mainz, 55128 Mainz, Germany}
\author{Teng Wu}
\affiliation{Johannes Gutenberg-Universit{\"a}t  Mainz, 55128 Mainz, Germany}
\author{Kasper Jensen}
\affiliation{Niels Bohr Institute, University of Copenhagen, Blegdamsvej 17, 2100 Copenhagen, Denmark}
\author{Yannick Dumeige}
\affiliation{CNRS, UMR 6082 FOTON, Enssat, 6 rue de Kerampont, CS 80518, 22305 Lannion cedex, France}
\author{Dmitry Budker}
\affiliation{Johannes Gutenberg-Universit{\"a}t  Mainz, 55128 Mainz, Germany}
\affiliation{Helmholtz Institut Mainz, 55099 Mainz, Germany}
\affiliation{Department of Physics, University of California, Berkeley, CA 94720-7300, USA}
\affiliation{Nuclear Science Division, Lawrence Berkeley National Laboratory, Berkeley, CA 94720, USA}



\date{\today}

\maketitle
\subsection{NV singlet state population}

We can calculate the NV state population from the increase in cavity losses under saturating green pump light.
The losses are related to the measurable quantities finesse ($\mathcal{F}$) and the cavity's IR light transmission on resonance to initial IR light ratio ($\frac{I_{\textrm{tr}}}{I_{\textrm{in}}}$).
Measuring the losses with ($A_{\textrm{2}}$) and without ($A_{\textrm{1}}$)  green pump light allows us to infer the losses due to NVs in the singlet state ($A_{\textrm{NV}}$). Finally, the increase in losses per roundtrip can be related to the NVs singlet population by the Beer-Lambert law. Assuming ideal loss-free mirrors finesse and transmission are given by:
\begin{equation}
\left(\frac{2\mathcal{F}}{\pi}\right)^2 = \frac{4\,\,\sqrt[]{R_1R_2(1 - A)}}{(1-\sqrt[]{R_1R_2(1-A))}^2}
\end{equation}
\begin{equation}
\frac{I_{\textrm{tr}}}{I_{\textrm{in}}} = \frac{4(1-R_1)(1-R_2)(1-A)}{(2-R_1-R_2+A)^2}
\end{equation}
where $R_1$ the reflectivity of the diamond coating, R$_2$ is the reflectivity of the output mirror and $A$ represents the losses per roundtrip due to the diamond crystal.\\
\indent According to the company specification, $R_1$=98.5(5)\%. $R_2$ and $A_{\textrm{1}}$ can be inferred solving Eq.\,(S1-S2) with measurements of finesse and transmission. Without green pump light we determine $R_2$\,=\,99.2(8)\% and $A_{\textrm{2}}$\,=\,1.66(1)\%.
Assuming the same reflectivity and adding saturating green pump light we get a new value for losses $A_{\textrm{2}}$\,=\,3.09(3)\%. The difference in losses $\Delta A$\,=\,$A_{\textrm{NV}}$ can be attributed to absorption on the singlet transition. These losses can be related to NV population in the singlet state via the Beer-Lambert law.
The transmission of the diamond for losses $A_{\textrm{NV}}$ is given by:
\begin{equation}
1-A_{\textrm{NV}} = e^{-n\sigma l}
\end{equation}
Where n is the density of the NVs in the singlet state, $\sigma$ the absorption cross section of IR light on resonant with the singlet transition and $l$ the path length in the diamond. With a path length $l\,=\,2L_\textrm{d}$ (where $L_\textrm{d}$=\,0.39\,mm is the geometric length diamond) and $\sigma\,=\,3\times10^{-18}\,$cm$^2$ (Ref.\cite{Dumeige2013}) we calculate the density of NVs in the metastable state as $n$\,=\,0.68(1)\,ppm.

\subsection{Cavity finesse} 

\begin{figure}[H]
\centering
\includegraphics[width=\columnwidth]{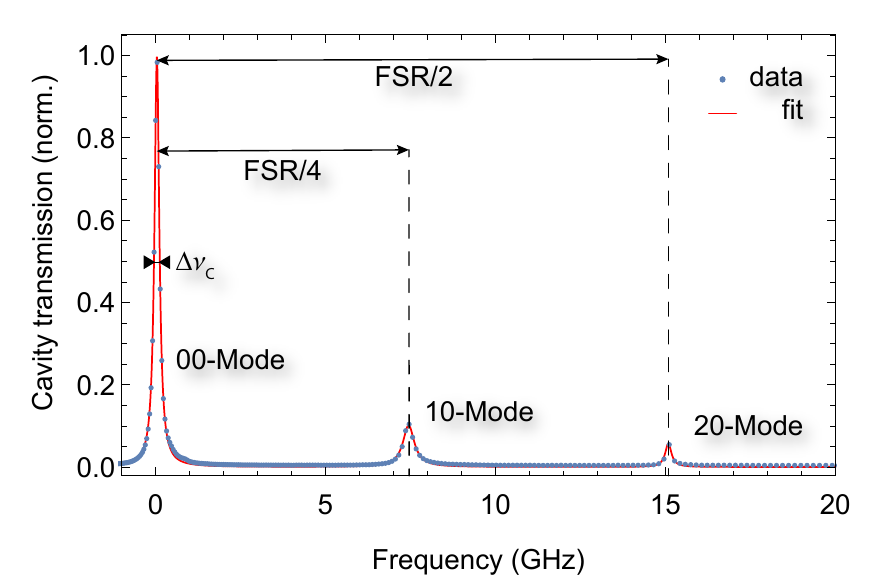}
\caption{\small{IR transmission spectrum of the cavity obtained by scanning the cavity length, in the absence of green light. Data are represented with dots \color{blue} (blue) \color{black} and the fit with a line \color{red} (red)\color{black}. }}
\label{fig:Sfig1}
\end{figure}

\indent Figure\,\ref{fig:Sfig1} shows the transmission of IR light while scanning the length of the cavity. From our measurements, after fitting we estimate the finesse ($\mathcal{F}$) of our cavity to be $\mathcal{F}$\,=\,160(4) [the FSR of the cavity is $\sim$\,30.1(2)\,GHz]. The finesse is measured in the absence of green pump light.\\

\subsection{Magnetometer Calibration} 
\raggedbottom
\begin{figure}[H]
\centering
\includegraphics[width=\columnwidth]{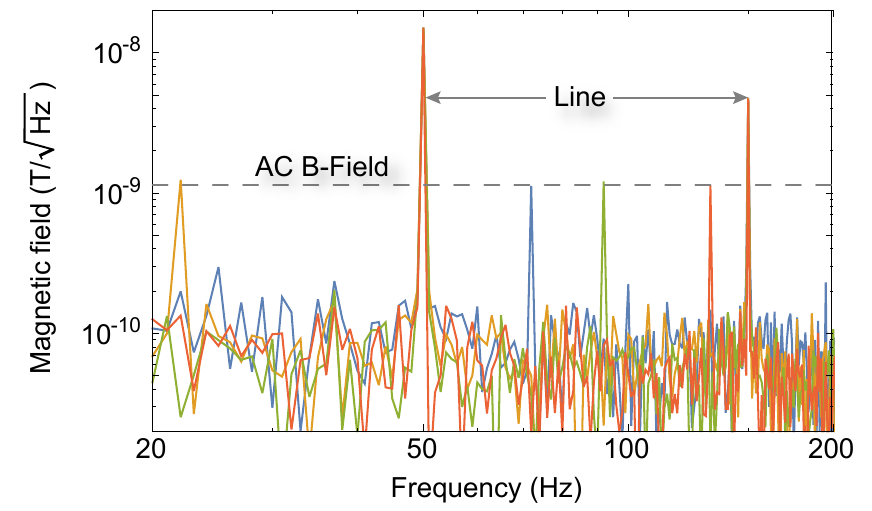}
\caption{Magnetic-field noise spectrum for different fields as a function of frequency. The test field is applied for different frequencies of 22\,Hz \color{amber} (amber)\color{black}, 72\,Hz \color{blue} (blue)\color{black}, 92\,Hz \color{green} (green) \color{black} and 132\,Hz \color{red} (red)\color{black}.}
\label{fig:Sfig2}
\end{figure}
The sensitivity of the sensor is verified by applying external magnetic fields of a given amplitude and frequency to it. Fig.\,\ref{fig:Sfig2} shows the noise spectrum of the magnetometer in response to a sine-wave-modulated magnetic field of 1\,nT RMS amplitude, at different frequencies.\\
\indent The field is provided by a multi-turn circular current coil. The radius of the coil is $r_{coil}\sim$2.5\,cm and its distance from the sensor is z$_{coil}\sim$1\,cm. The coil is connected in series with a 1\,k$\Omega$ resistor to decrease induction induced current. The amplitude of the produced magnetic field B$_{test}$ is then compared with the recorded magnetometer signal B$_{rec}$. B$_{test}$ can be calculated as:
\begin{align*}
     B_{test} =\frac{\mu_{0}\,N_{turns}\,I_{coil}\,r^2_{coil}}{2[z^2_{coil}+r^2_{coil}]^{3/2}}.
  \end{align*}
where $\mu_0$\,=\,4$\pi\times10^{-7}$ N/A$^2$ is the vacuum permeability, $N_{turns}$\,=\,11 are the number of turns of the coil and $I_{coil}$\,=\,6.5\,$\mu$A the current flowing through the coil. Our measured $B_{rec}$ is consistent with $B_{test}$ within a 2\% error.

\subsection{Photon shot-noise behavior} 
\begin{figure}[H]
\centering
\includegraphics[width=\columnwidth]{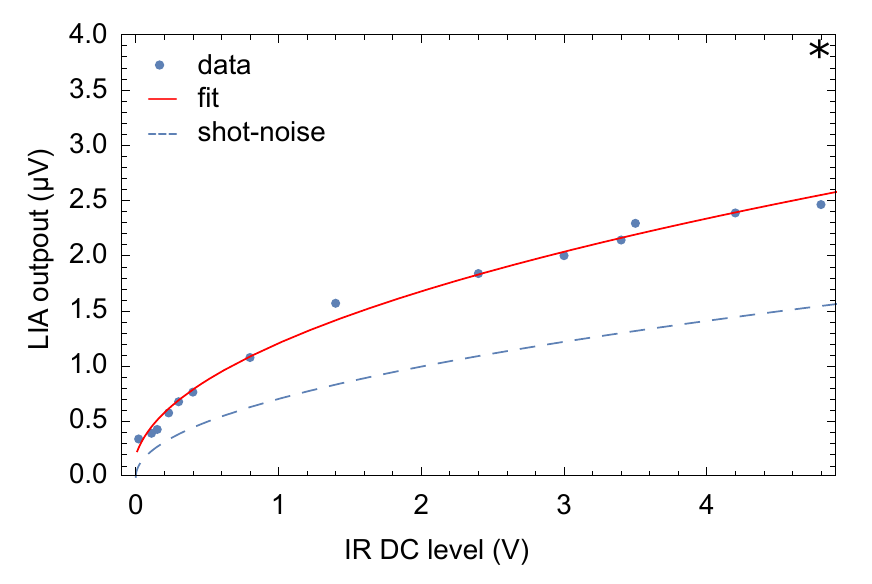}
\caption{Measured and calculated RMS noise of cavity transmission. Data show in dots \color{blue} (blue)\color{black}, are fitted with a line  \color{red} (red)\,\color{black}. Theoretical calculation for the shot noise is presented with dashed lines and an asterix symbol (*) is placed to signify the noise level under normal operation.}
\label{fig:Sfig3}
\end{figure}

	To confirm the photon shot noise behavior of our sensor, in the absence of MWs, we record the output of the LIA ($S_\textrm{IR}$) as we detect transmission ($I_\textrm{tr}$) for a wide range of IR light powers. We then fit the data with a curve $S_\textrm{IR} = \sqrt[]{a^2+b^2 I_\textrm{tr}+c^2 I_\textrm{tr}^2}$, where $a$ signifies laser power independent noise sources (e.g. electronic noise), $b$ relates to IR light shot-noise and $c$ to IR intensity noise sources.\\
\indent Figure\,\ref{fig:Sfig3} shows the above measurements in dots (blue) at 100\,mW of green light. The data is fitted (red) giving fit values $a$\,=\,0.23(8), $b$\,=\,1.16(6) and $c$\,=\,2$\times$10$^{-3}$(2) which indicates shot-noise dominated behavior for IR DC levels above $\sim$\,0.2\,V. The theoretically calculated shot noise is represented by the dashed line. Finally, the asterisk (*) is placed in Fig.\,\ref{fig:Sfig3} to signify the average noise floor between 60-90\,Hz under normal operation corresponding to 400\,mW of green pump light and a MW field applied to the NVs. The noise floor thus includes environmental and MW-related noise.
\bibliography{literature}